\documentclass[final,5p,times,twocolumn,number]{elsarticle}
\usepackage{xcolor}
\usepackage{amssymb}
\usepackage{amsmath}

\usepackage{lineno}
\usepackage[version=4]{mhchem}
\usepackage[hidelinks]{hyperref} 

\newcommand{\muB}{\ensuremath{\mu_{\mathrm B}}}

\newcommand{\sNN}{\ensuremath{\sqrt{s_\text{NN}}}}

\newcommand{\dNdy}{\ensuremath{\mathrm d N/\mathrm d y}}
\newcommand{\dNchdeta}{\ensuremath{\mathrm d N_{\text{ch}}/\mathrm d\eta}}
\newcommand{\Vdy}{\ensuremath{V_{\Delta y=1}}}

\newcommand{\hypertriton}{\ce{^{3}_{\Lambda}H}} 
\newcommand{\hyperHfour}{\ce{^{4}_{\Lambda}H}}
\newcommand{\hyperHefour}{\ce{^4_\Lambda He}} 
\newcommand{\helion}{\ce{^3He}} 
\newcommand{\hefour}{\ce{^4He}}
\newcommand{\triton}{\ce{^3H}}

\journal{Physics Letters B}

\begin{document}

\begin{frontmatter}

\title{Hadronization into Nuclei: Does Size Matter?}

\author[first]{Anton Andronic}
\author[second,third,fourth]{Peter Braun-Munzinger}
\author[second,fifth]{Hjalmar Brunßen\corref{cor1}}
\ead{hjalmar.brunssen@cern.ch}
\cortext[cor1]{Corresponding author}
\author[second,fourth]{Johanna Stachel}

\affiliation[first]{organization={Institut für Kernphysik, Universität Münster},
            city={Münster},
            country={Germany}}
\affiliation[second]{organization={Physikalisches Institut, Universität Heidelberg},
            city={Heidelberg},
            country={Germany}}

\affiliation[third]{organization={ExtreMe Matter Institute EMMI, GSI Helmholtzzentrum für Schwerionenforschung},            city={Darmstadt},
            country={Germany}}

\affiliation[fourth]{organization={Institute of Particle Physics and Key Laboratory of Quark and Lepton Physics (MOE),
Central China Normal University},
            city={Wuhan},
            country={China}}

\affiliation[fifth]{organization={School of Physics and Astronomy, University of Birmingham},
            city={Birmingham},
            country={United Kingdom}}
            
\begin{abstract}
We investigate an extension of the statistical hadronization model (SHM) that takes (hyper)nucleus sizes into account. The spatial extent of nuclear wave functions relative to the fireball volume is used to calculate a size correction, which describes the suppression with respect to a point-like treatment of the (hyper)nucleus at hadronization. For several species of light nuclei (d, \triton, \helion) and for the hypernucleus \hypertriton, the predicted effects of the size correction on SHM yields are compared to yield measurements in pp, p--Pb, and Pb--Pb collisions in the ALICE experiment. The experimental data suggest that this size correction, based on applying final-state wave functions at chemical freeze-out, does not give a consistent description of (hyper)nucleus production in nuclear and hadronic collisions. By contrast, the SHM without a nuclear-size correction provides a substantially better description of the data. This result lends further support to the interpretation that such nuclear states are formed from initially compact (multi-quark) configurations which after hadronization expand to their final-state wave functions.
\end{abstract}


\end{frontmatter}

\section{Introduction}
\label{sec:introduction}
In high-energy heavy-ion collisions, a wide variety of hadrons, light nuclei, and hypernuclei are produced. The production mechanism of these composite states is still actively debated, with the statistical hadronization model (SHM) \cite{Hagedorn:1965st,Andronic:2026cdu} and coalescence approaches~\cite{Butler:1963pp,Kapusta:1980zz} representing the two main conceptual frameworks. In the SHM, the abundances of hadrons in a heavy-ion collision are described in a statistical ensemble~\cite{Braun-Munzinger:2003pwq,Andronic:2017pug}. Their densities are determined by their mass, quantum numbers, and hadro-chemical freeze-out parameters: in the grand-canonical ensemble these parameters are the temperature $T$ and the baryo-chemical potential \muB. To obtain rapidity densities, one needs in addition the volume $V$, corresponding e.g., to one unit of rapidity.
It has been demonstrated in previous studies that experimental rapidity densities are excellently described by the SHM over nine orders of magnitude in abundance~\cite{Andronic:2017pug,Andronic:2005yp,Andronic:2026cdu}. A remarkable feature is that the SHM describes not only the production of hadrons but likewise the yields of (anti)nuclei (d, \triton, \helion{}, \hefour) and even of the hypertriton \hypertriton{}. This is non-trivial given their large sizes and tiny binding energies. For example, the deuteron has a binding energy $E_B = 2.2\,\mathrm{MeV}$~\cite{Wang:2021xhn}, and \hypertriton{} has a $\Lambda$ separation energy $B_\Lambda=0.102\pm0.092\,\mathrm{MeV}$~\cite{ALICE:2022sco}, both far below the chemical freeze-out temperature $T\approx100-150$~MeV.

While the grand-canonical SHM successfully describes yields of hadrons and (hyper)nuclei in Pb--Pb collisions, smaller collision systems (pp, p--Pb) require treatment in a canonical ensemble, where quantum numbers are exactly conserved within a correlation volume $V_\mathrm{c}$~\cite{Hamieh:2000tk,Braun-Munzinger:2003pwq,Beutler:2010cha}. This results in a canonical suppression factor relative to the grand-canonical expectation. It can be evaluated by a ratio of modified Bessel functions of the first kind. In the case of strangeness, the suppression can be written as
\begin{equation}
	n_s^{C} \simeq n_s^{\mathrm{GC}} \frac{I_s\left(2\sqrt{N_{S=+1}^{\mathrm{th}}\,N_{S=-1}^{\mathrm{th}}}\right)}{I_0\left(2\sqrt{N_{S=+1}^{\mathrm{th}}\,N_{S=-1}^{\mathrm{th}}}\right)}\,,
\end{equation}
where the strangeness  $s$ of the particle species of interest enters as the order of the modified Bessel function and the argument is determined by the thermal number $N_{S=+1}^{\mathrm{th}}$ of hadrons with strangeness quantum number $S=+1$ inside $V_\mathrm{c}$ and its $S=-1$ counterpart. 
The exact conservation of baryon number can be implemented analogously~\cite{Beutler:2010cha,Sharma:2022poi,Braun-Munzinger:2023gsd}.

While various SHM predictions for (hyper)nuclei\,\cite{Andronic:2017pug,Vovchenko:2019kes,Sharma:2022poi} agree well with data, other approaches suggest that a late production, around the kinetic freeze-out, also describes the data \,\cite{Xu:2018jff,Oliinychenko:2018ugs,Vovchenko:2019aoz,Sun:2022xjr,Knoll:2024gaf}, supporting a coalescence-type mechanism\,\cite{Mrowczynski:2020ugu}.
The production of loosely bound states was recently addressed in a novel approach\,\cite{Braaten:2024cke}, while a direct critique of an initially compact precursor quark-gluon bag\,\cite{Andronic:2017pug} was put forward\,\cite{Cohen:2024wgs}.

In this paper, the production of nuclei and hypernuclei is studied using an extension of the SHM that takes into account their spatial extent, exploring a recent proposal by Müller~\cite{Muller:2022uuv}. This approach is compared to experimental data from pp, p--Pb, and Pb--Pb collisions at the ALICE experiment.
\section{Correction for nuclear size}
\label{sec:sizecorr}
The hypertriton is so loosely bound that its size ($\Lambda$--d root-mean-square (rms) distance $\sim\!10\,\mathrm{fm}$~\cite{Braun-Munzinger:2018hat}) is comparable to the fireball radius in central Pb--Pb collisions (see below). This raises an apparent paradox: as the wave function extends beyond the fireball volume, there is a sizable probability that one of the constituents lies outside the fireball.
As described in Ref.~\cite{Muller:2022uuv}, this paradox is resolved by introducing a size correction that requires all hypertriton constituents to lie within the fireball volume.\\
In the simplest case of a two-body nucleus, e.g., the deuteron, or the hypertriton treated as a $\Lambda$--d bound state, the size correction $\mathcal{S}_i$ is given by
\footnote{In Müller's paper~\cite{Muller:2022uuv}, the size correction is denoted $P(a)/P(0)$.}
\begin{align}
	\mathcal{S}_i &= \frac{1}{{V_{\Delta y=1}}}\:\underset{V_{\Delta y=1}}{\int}\mathrm{d}^3 x_1\underset{V_{\Delta y=1}}{\int} \mathrm{d}^3 x_2\;
		\bigl|\psi_i(\mathbf{x}_1 - \mathbf{x}_2)\bigr|^2.
\end{align}
The distance distribution enters in the form of the spatial wave function $\psi(\mathbf{x}_1 - \mathbf{x}_2)$, where $\mathbf{x_1}$ and $\mathbf{x_2}$ are the position vectors of both constituents.
This correction factor modifies the SHM yields of (hyper)nuclei as 
\begin{align}
	\frac{\mathrm{d}N}{\mathrm{d}y}\bigg|_{\Delta y = 1} &= \mathcal{S}_i \;n_i(T, \mathbf{\muB})\; V_{\Delta y=1}.\label{eq:sizecorr}
\end{align}
The size correction $\mathcal{S}_i$ expresses the probability of finding all constituents of a given nucleus within the fireball volume $V_{\Delta y=1}$, considered here for one unit of rapidity at midrapidity. It approaches $\mathcal{S}_i=1$ in the limit of nuclei that are compact, whereas for (hyper)nuclei with a spatial extent comparable to the fireball size, a substantial suppression, $\mathcal{S}_i<1$, is expected.\\
The size correction approach can also be applied to $A\geq3$ (hyper)nuclei. As a halo nucleus~\cite{Gibson:2001kf,Hammer:2001ng,Braun-Munzinger:2018hat}, the hypertriton can be described as a $\Lambda$ that is very loosely bound to a deuteron core. The spatial wave function can therefore be factorized into a part describing the $\Lambda$--d separation, $\psi_{\Lambda}$, and the deuteron wave function $\psi_{\mathrm{d}}$, describing the distance between the neutron and the proton. For brevity, we denote $\Vdy$ by $V$ in the following. The size correction for the hypertriton is given by
\begin{align}
 \mathcal{S_{\hypertriton}}= &\frac{8}{{V}}\underset{V}{\int}\mathrm{d}^3 x_\mathrm{d} \underset{V}{\int}\mathrm{d}^3 x_1\;\underset{V}{\int}\mathrm{d}^3 x_\Lambda\;\bigl|\psi_{\Lambda}(\mathbf x_{\Lambda}-\mathbf x_\mathrm{d})\bigr|^{2}\notag\\
    &\times
	\bigl|\psi_\mathrm{d}\bigl(\underbrace{2(\mathbf{x}_1 - \mathbf{x}_\mathrm{d})}_{\mathbf{x}_1-\mathbf{x}_2}\bigr)\bigr|^2\,\Theta_{V}\bigl(\underbrace{2\mathbf{x}_\mathrm{d} -\mathbf{x}_1}_{\mathbf{x}_2}\bigr),\label{eq:hyptsizecorr}
\end{align}
where $\mathbf{x}_\mathrm{d}$ and $\mathbf{x}_\Lambda$ denote the deuteron center of mass and the $\Lambda$ position, and the Heaviside function $\Theta_{V}\bigl(\mathbf{x}_2\bigr)$ ensures that the second nucleon also lies within the fireball. The factor 8 comes from the Jacobian of the change of variables $(\mathbf{x}_1,\mathbf{x}_2)\rightarrow(\mathbf{x}_1,\mathbf{x}_\mathrm{d})$.

For \helion{} and \triton, we use nucleon distributions relative to their center of mass $\mathbf{x}_{\helion}$ in the form of normalized one-particle density distributions $\rho_\mathrm{p}$ and $\rho_\mathrm{n}$. The size correction for, e.g., \helion{} is calculated as 
\begin{align}
	\mathcal{S_{\helion}}
	&=
	\frac{\mathcal{N}}{{V}}\underset{V}{\int}\mathrm{d}^3 x_{\helion{}}
    \underset{V}{\int} \mathrm d^{3} x_{1}
	\underset{V}{\int} \mathrm d^{3} x_{2}\,
	\rho_\mathrm{p}\bigl(\lvert\mathbf x_{1}-\mathbf x_{\helion}\rvert\bigr)\,\notag\\
	&\times
	\rho_\mathrm{p}\bigl(\lvert\mathbf x_{2}-\mathbf x_{\helion}\rvert\bigr)\,
	\rho_\mathrm{n}\bigl(\underbrace{\lvert2\,\mathbf x_{\helion}-\mathbf x_{1}-\mathbf x_{2}\rvert}_{\lvert\mathbf x_3-\mathbf x_{\helion}\rvert}\bigr)\,\notag\\
	&\times\Theta_{V}\bigl(\underbrace{3\,\mathbf x_{\helion}-\mathbf x_{1}-\mathbf x_{2}}_{\mathbf x_3}\bigr), \label{eq:sizecorrHe3}
\end{align}
where the normalization factor $\mathcal{N}$ ensures that the size correction converges to 1 in the large-volume limit. The analogous formula has been applied to \triton{} and can be extended to any $A\geq3$ nucleus. 

\section {Sizes of fireball and nuclei}
\label{sec:volumewavefunctions}
While the midrapidity fireball volume \Vdy{} can be determined within the statistical hadronization model, evaluating the size correction also requires the fireball geometry to be specified. In the following, we assume a cylindrical fireball. Its length is chosen based on the freeze-out proper time $\tau_\mathrm{f}$ extracted from two-pion Bose-Einstein correlations. Assuming longitudinal flow according to the Bjorken model~\cite{Bjorken:1982qr}, $\tau_\mathrm{f}$ is translated into a longitudinal length scale using the relation
\begin{align}
    z=\tau_\mathrm{f} \sinh y,
\end{align}
which relates the longitudinal coordinate $z$ to the space-time rapidity $y$.
The freeze-out proper time $\tau_\text{f}=10.4\pm0.5~\mathrm{fm}/c$~\cite{ALICE:2011dyt} measured in central Pb--Pb collisions at $\sNN=2.76\,\mathrm{TeV}$ corresponds to a longitudinal length $L=10.9~\mathrm{fm}$ of the central unit of rapidity. Together with the midrapidity volume $\Vdy=4175\pm380\,\mathrm{fm^3}$~\cite{Andronic:2021dkw} from the SHM, this fixes the transverse radius of the cylinder to $R=11.0~\mathrm{fm}$. This aspect ratio between length and transverse radius is applied to all collision systems, which is supported by the observed scaling $\tau_\mathrm{f}\propto(\dNchdeta)^{1/3}$~\cite{ALICE:2011dyt}.
It was verified that the size correction depends primarily on the fireball volume \Vdy{} and only weakly on the assumed shape and relative dimensions of the fireball for reasonable variations.\\

\begin{figure}[htbp]
	\centering 
	\includegraphics[width=\linewidth]{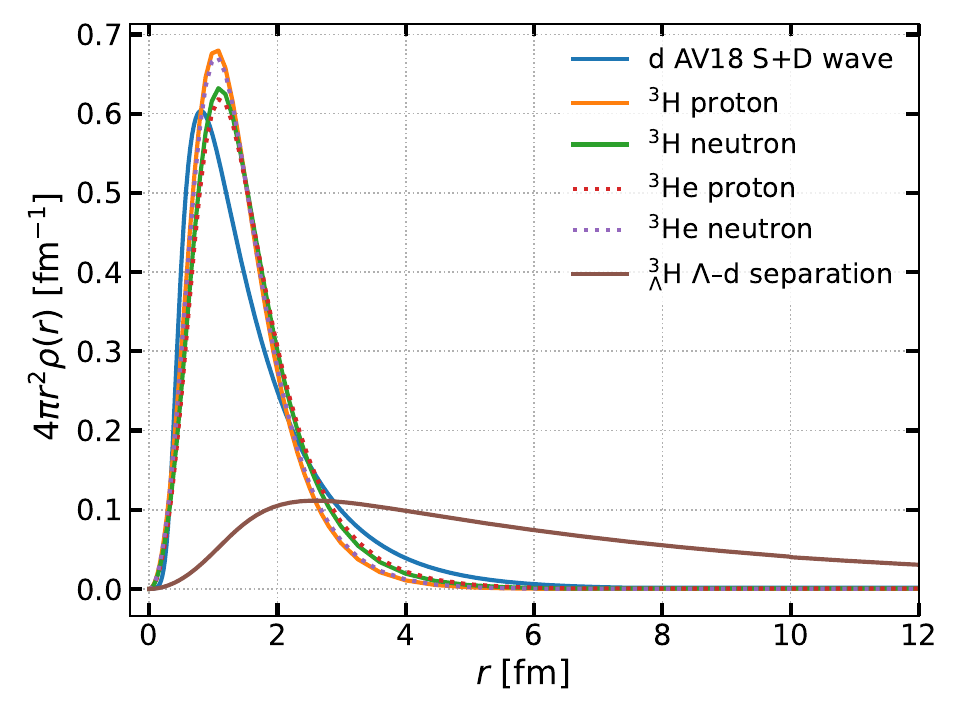}
	\caption{Radial probability density distributions for the studied (hyper)nuclei. The proton and neutron probability density distributions, shown as a function of the nucleon distance from the center of mass, closely follow the expected approximate isospin symmetry between $^3\mathrm{H}$ (solid) and $^3\mathrm{He}$ (dotted). To allow a direct comparison, the Argonne $v_{18}$ deuteron wave function is plotted analogously, with $r$ denoting half the n--p separation. The \hypertriton{} curve represents the distance distribution of the $\Lambda$ hyperon from the deuteron core and illustrates its long tail ~\cite{Braun-Munzinger:2018hat} extending even far beyond $10\,\mathrm{fm}$.} 
	\label{fig:distributions}
\end{figure}

For the deuteron, the wave function is based on the Argonne $v_{18}$ nucleon-nucleon potential~\cite{Wiringa:1994wb}, obtained by fitting NN scattering data and the deuteron binding energy. The spatial wave function~\cite{ANLdeutWF} contains an S- (94\%) and D-wave (6\%) component and has an rms matter radius of $\sqrt{\langle r_\text{m}^2\rangle}=1.94~\mathrm{fm}$.

The internal structure of  the hypertriton is described using radial
correlation densities obtained from chiral effective field theory (EFT) calculations based on semi-local momentum-space (SMS)–regularized potentials~\cite{Haidenbauer:2025zrr,Nogga2003,Nogga:PrivateCommunication}. The resulting
$\Lambda$--d and n--p distance distributions are used to calculate the size correction. The $\Lambda$--d distribution has an rms separation of $\sqrt{\langle{r_{\Lambda\mathrm{d}}}^2\rangle}=11.1\,\mathrm{fm}$, whereas the proton and neutron form a much more compact, deuteron-like core. The calculated internal n--p distribution closely resembles
that obtained from the Argonne $v_{18}$ wave function of a free deuteron. The
pronounced separation of length scales between the compact n--p core and the spatially extended
$\Lambda$ distribution demonstrates the halo structure of the
\hypertriton{}.

The size correction for \helion{} and \triton{} is calculated based on one-particle density distributions $\rho_\mathrm{p}$ and $\rho_\mathrm{n}$ of the proton and neutron inside the nuclei, which are normalized as $\int_0^\infty 4\pi r^2 \rho(r) dr=1$. It should be noted that these do not contain the full information of the wave function as they omit correlations between the individual nucleons. However, we believe that one-body densities are sufficient for the purpose of this work because microscopic details are largely averaged out in the size-correction approach. For \helion{} and \triton{}, the one-body densities were obtained from chiral EFT calculations based on SMS–regularized potentials~\cite{Nogga2003,Nogga:PrivateCommunication}. A comparison of the different normalized radial probability distributions is shown in Fig.~\ref{fig:distributions}.

\section{Results}
\begin{figure}[htbp]
	\centering 
	\includegraphics[width=\linewidth]{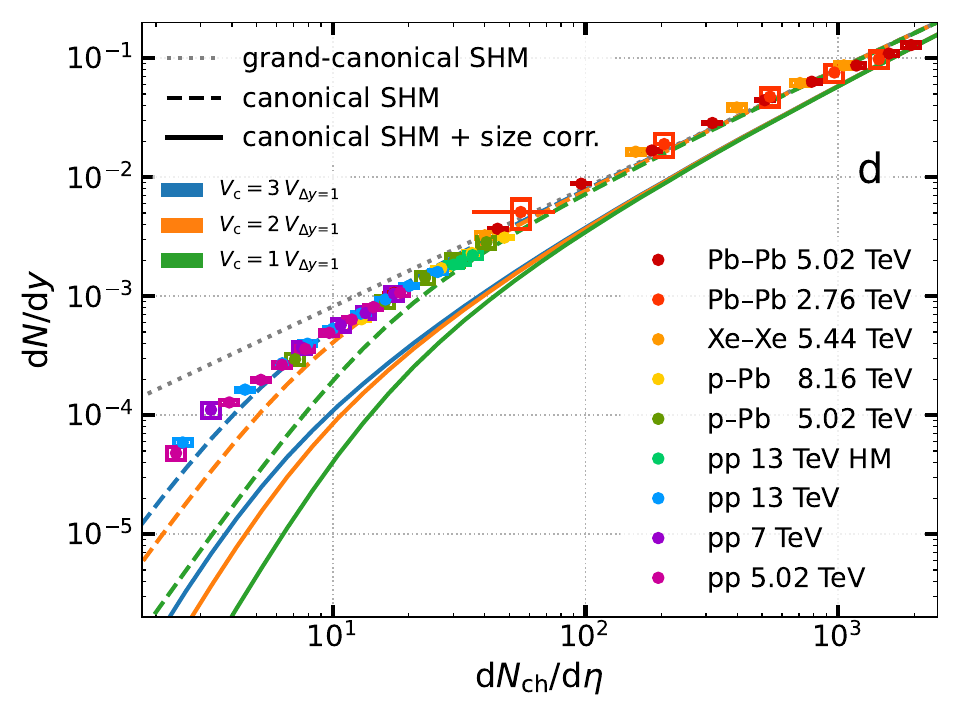}
	\caption{Deuteron rapidity densities at midrapidity depicted for a wide range of system sizes characterized by the charged-particle multiplicity \dNchdeta. The gray dotted line shows the grand-canonical SHM values~\cite{Andronic:2017pug}, dashed lines indicate the canonical SHM results for different values of the canonical volume, and results for canonical SHM with size correction are shown as solid lines. Line colors indicate the canonical volume $V_\text{c}$, as specified in the legend. Measurements from the ALICE collaboration for Pb--Pb~\cite{ALICE:2022veq, ALICE:2015wav}, Xe--Xe~\cite{ALICE:2024say}, p--Pb~\cite{ALICE:2019bnp, ALICE:2022pbb}, and pp~\cite{ALICE:2021mfm,ALICE:2020foi, ALICE:2019dgz, ALICE:2021ovi} collisions are shown with statistical and systematic uncertainties as vertical bars and open boxes, respectively. For each of these collision systems, multiple centrality or multiplicity classes are presented. Throughout this work, the average of nucleus and antinucleus yields is shown whenever both are available.} 
	\label{fig:deuteron}
\end{figure}

In Fig.~\ref{fig:deuteron}, the results for the rapidity density \dNdy{} of deuterons are shown as a function of charged-particle multiplicity \dNchdeta. In addition to measurements for various collision systems (Pb--Pb, Xe--Xe, p--Pb, and pp), the corresponding results from the grand-canonical SHM are shown along with curves containing the canonical suppression as well as the canonical suppression and size correction. For the calculations we use the same thermal parameters for all collision systems and, to obtain rapidity densities, scale the volume obtained from data comparison in central Pb--Pb collisions ($\Vdy=4997\,\mathrm{fm^3}$ at $\sNN=5.02\;\mathrm{TeV}$~\cite{Andronic:2021erx}) proportional to the charged-particle multiplicity, following experimental systematics. The canonical suppression is computed for 3 different values for the canonical volume over which the quantum number (here baryon number) is conserved. We choose the canonical volume corresponding to 1, 2, and 3 units of rapidity. Results for second-order cumulants of net-proton distributions favor the larger value~\cite{ALICE:2022xpf}. For charged-particle multiplicity densities below 50, the effect of the canonical suppression becomes increasingly visible, and the data follow this expected suppression well. The size correction adds an additional significant suppression of nearly one order of magnitude for the smallest systems, and the corresponding calculations fall significantly below the experimental data, especially for intermediate and small fireball volumes (\dNchdeta). A very large dynamic range of 3 orders of magnitude in \dNchdeta{} and a corresponding drop in deuteron multiplicity by about a factor of 3000 are reproduced quantitatively by implementing canonical thermodynamics.

\begin{figure}[htbp]
\centering 
\includegraphics[width=\linewidth]{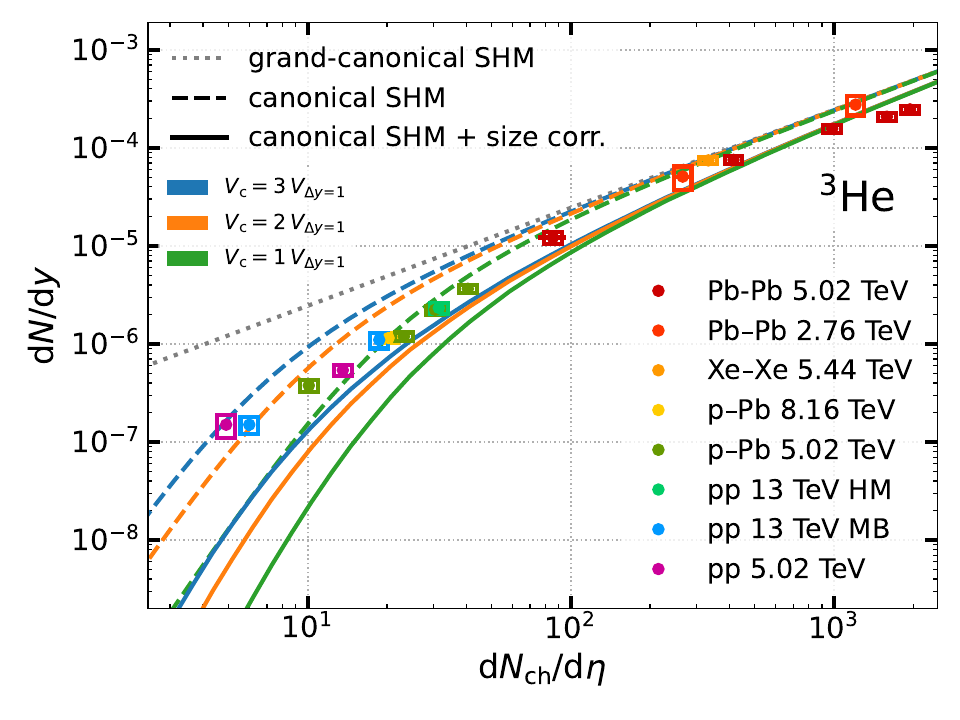}	
\caption{Multiplicity dependence of the \helion{}  rapidity density at midrapidity. ALICE measurements for Pb--Pb~\cite{ALICE:2022veq,ALICE:2015wav}, Xe--Xe~\cite{ALICE:2024say}, p--Pb~\cite{ALICE:2022pbb,ALICE:2019fee}, and pp collisions ~\cite{ALICE:2021mfm,ALICE:2021ovi} are depicted and compared to results for the same set of calculations (see text). The line styles and colors for the SHM calculations follow the same notation as described in Fig.~\ref{fig:deuteron}. } 
\label{fig:helion}
\end{figure}

\begin{figure}[htbp]
\centering 
\includegraphics[width=\linewidth]{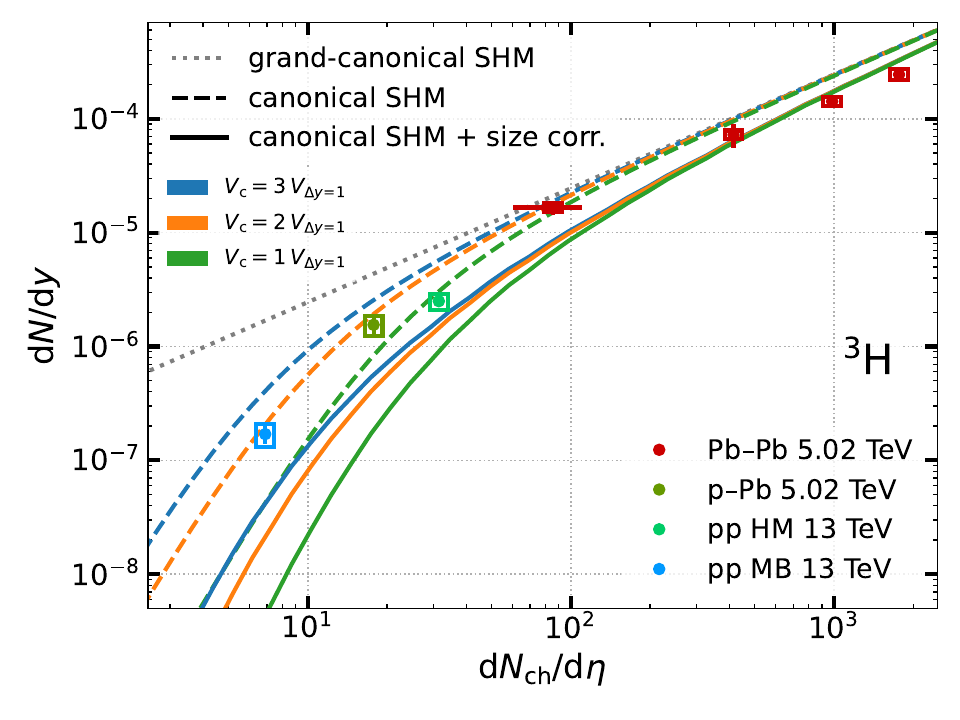}	
\caption{Multiplicity dependence of the \triton{} rapidity density at midrapidity. ALICE measurements for Pb--Pb~\cite{ALICE:2022veq}, p--Pb~\cite{ALICE:2019fee}, and pp collisions~\cite{ALICE:2021mfm} are shown and compared to results for the same set of calculations (see text). The line styles and colors for the SHM calculations are the same as in Fig.~\ref{fig:deuteron}.} 
\label{fig:triton}
\end{figure}

The charged-particle multiplicity dependence of the \helion{} and \triton{} rapidity densities is shown in Figs.~\ref{fig:helion} and~\ref{fig:triton}, respectively. Again, experimental data from the ALICE collaboration are compared to the results for the different statistical ensembles (see the description of Fig.~\ref{fig:deuteron}). The data are more sparse, in particular for \triton, which is much more difficult to identify in the important momentum range using a time projection chamber (TPC). The Pb--Pb measurements at $\sNN=5.02\,\mathrm{TeV}$ are somewhat below the yields expected from statistical hadronization and may be closer to the results from the size-corrected calculations. 
But the experimental charged-particle multiplicity dependence of the yields of mass-3 nuclei follows more closely the drop of SHM results with canonical suppression only and the data clearly exceed the results of the size-suppressed calculations, in particular for small multiplicities where the effects are most pronounced. This is in line with the observation for the deuteron.

For the light nuclei (d, \helion, \triton), the prediction of a large size correction in small collision systems (pp, p--Pb) due to the spatial extent of the nuclear wave function is not supported by the measurements. The expected effect of the nuclear size is even stronger for the hypertriton than for the non-strange light nuclei due to its very large spatial extent (see Fig.~\ref{fig:distributions}), making it a particularly sensitive probe of a potential nuclear size correction effect.

\begin{figure}[htbp]
	\centering 
	\includegraphics[width=\linewidth]{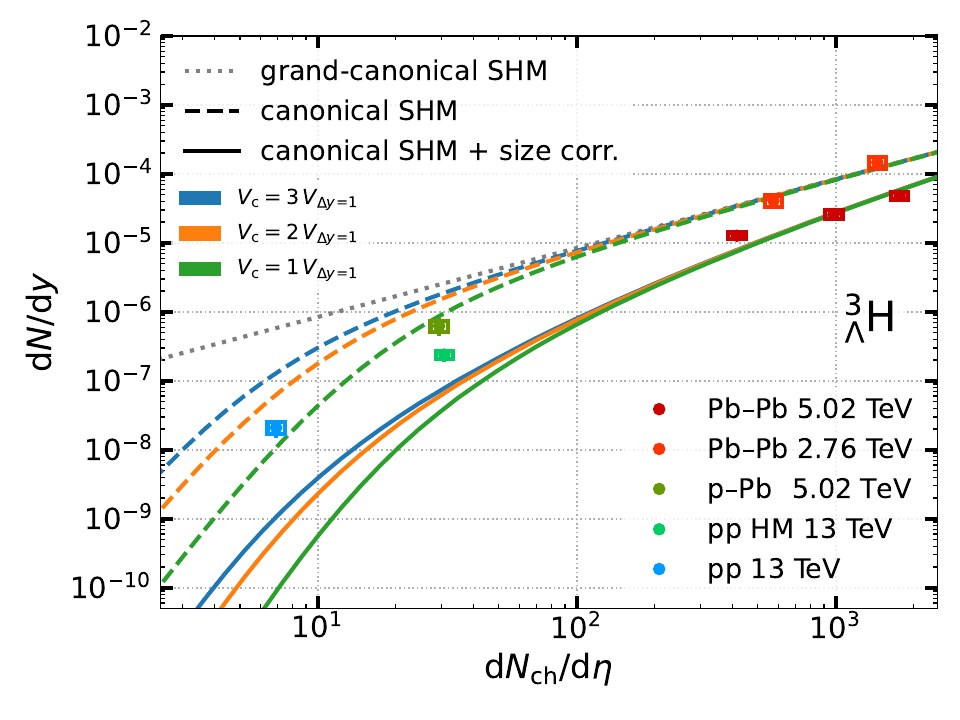}	
	\caption{Rapidity density   of \hypertriton{} shown as a function of charged-particle multiplicity \dNchdeta. ALICE measurements for Pb--Pb~\cite{ALICE:2024koa,ALICE:2015oer}, p--Pb~\cite{ALICE:2021puh}, and pp~\cite{ALICE:2026tqr} collisions are shown  and compared to results for the same set of calculations (see text). The line styles and colors for the SHM calculations are the same as in Fig.~\ref{fig:deuteron}.} 
	\label{fig:hypertriton}
\end{figure}

In Fig.~\ref{fig:hypertriton}, the multiplicity dependence of the hypertriton rapidity density at midrapidity is shown. The data situation is more complex here. The data points for Pb--Pb collisions at $\sNN=2.76~\mathrm{TeV}$ and p--Pb collisions at $\sNN=5.02~\mathrm{TeV}$ are fully consistent with the SHM including canonical suppression. In contrast, the Pb--Pb data at $\sNN=5.02~\mathrm{TeV}$ are a factor 3$\pm$1 lower than the $\sNN=2.76~\mathrm{TeV}$ measurement. The difference has not been explained. The absolute values for central Pb--Pb collisions are in agreement with the size-suppressed calculation. In this context we note that the considerable uncertainty in the measured $\Lambda$ separation energy for the hypertriton  is not yet considered in the hypertriton yield determination.  In view of the unclear data situation in Pb--Pb collisions it pays to concentrate on the multiplicity (i.e., system size) dependence. As for the other nuclei, the fall-off for the hypertriton is significantly less than predicted by the size-corrected SHM (for minimum bias pp collisions by more than one order of magnitude).

A similar pattern is visible in the yield ratio \hypertriton/$\Lambda$ that can be found in Fig.~\ref{fig:hypertritonoverLambda}.
For comparison, the multiplicity dependence of the $\Lambda$ yield is shown in \ref{fig:suppLambda}.

It is worth noting that interaction effects are not yet considered for the calculations of any nuclear production yields including \hypertriton, which could be quantified by calculations in the S-matrix approach. For the proton~\cite{Andronic:2018qqt,Cleymans:2020fsc} and even more so for the $\Lambda$ hyperon~\cite{Cleymans:2020fsc}, S-matrix calculations have been shown to modify thermal yields substantially. For the \hypertriton, the effect of interactions has not been quantified yet and could potentially have a significant impact on its predicted yield.\\
In addition to the (hyper)nuclei presented here, rapidity densities of \hefour~\cite{ALICE:2017jmf,ALICE:2023qyl} and the hypernuclei \hyperHfour{} and \hyperHefour~\cite{ALICE:2024djx} have been measured in Pb--Pb collisions by ALICE. For all three $A=4$ (hyper)nuclei, excellent agreement with the SHM predictions was found.

\section{Conclusions}
\label{sec:conclusion}
In this paper, we have studied an extension of the SHM that accounts for the size of nuclear wave functions. In this model, a size correction leading to a strong suppression is implemented for all nuclei and hypernuclei for which experimental data from the CERN LHC on yields over a large range of charged-particle multiplicity densities or fireball sizes exist. At smaller \dNchdeta, a canonical formulation of the SHM needs to be implemented before deciding on the necessity of a size correction. For all light nuclei studied (d, \helion, \triton), the statistical hadronization model with canonical suppression achieved a significantly better description without considering the nuclear size. The hypertriton, the most loosely bound of all (hyper)nuclei observed in heavy-ion collisions and therefore  the one with the most extended wave function, should be particularly sensitive.
While the data situation is more complex, the multiplicity dependence of hypertriton yields supports the same conclusion: the nuclear size correction leads to predicted yields which decrease much more steeply with decreasing multiplicity than the experimental data. This puts into question a formation scenario for nuclei where nucleon distances are characterized by the asymptotic nuclear wave functions. Such a result may not = too surprising, since the formation of a strongly bound object requires separations within the range of the strong interaction. A static $\Lambda$ hyperon placed at a distance of 10 fm from a deuteron will not form a bound hypertriton. As already pointed out previously~\cite{Andronic:2017pug}, the good agreement between ALICE data and the statistical hadronization model can be interpreted as indication of the formation of nuclei and hypernuclei as compact multi-quark objects with small virtuality and the quantum numbers of the final-state (hyper)nuclei. The quantum-mechanical wave function requires time to be formed. For example, for the deuteron with a binding energy of 2.2 MeV, the characteristic time scale $\hbar/E_\mathrm{B}$ is of order 100 fm/$c$, i.e., well beyond the decoupling time of the nuclear fireball, suggesting that the final-state nuclear wave function is established considerably after hadronization.

\appendix

\section{Supplementary material}
\label{sec:suppl}
\setcounter{figure}{0}
\renewcommand*{\theHfigure}{appendix.\Alph{section}.\arabic{figure}}
The multiplicity dependence of the hypertriton-to-$\Lambda$ yield ratio and the $\Lambda$ yield is shown in Figs.~\ref{fig:hypertritonoverLambda} and \ref{fig:suppLambda}, respectively.

\begin{figure}[htbp]
	\centering \includegraphics[width=\linewidth]{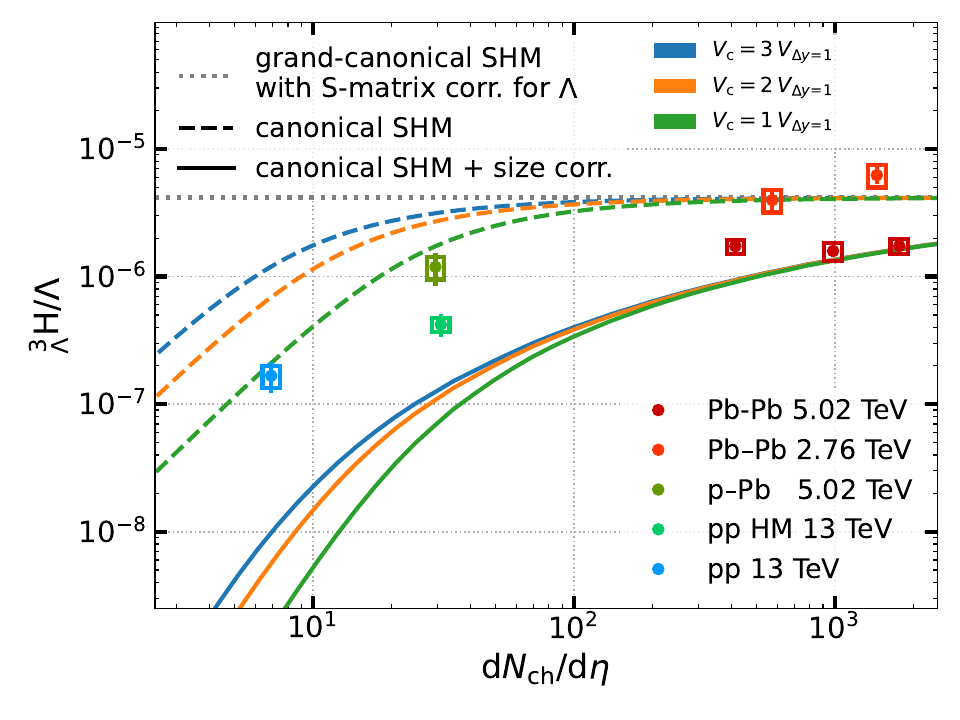}	
	\caption{Yield ratio of \hypertriton{} to $\Lambda$ as a function of charged-particle multiplicity \dNchdeta. The same line styles as in Fig.~\ref{fig:deuteron} are used. The S-matrix correction, which  accounts for interactions among hadrons in the hadron resonance gas, is applied to the SHM yield of $\Lambda$~\cite{Cleymans:2020fsc}. ALICE measurements for Pb--Pb~\cite{ALICE:2024koa,ALICE:2025cqy,ALICE:2015oer,ALICE:2013cdo}, p--Pb~\cite{ALICE:2021puh}, and pp~\cite{ALICE:2026tqr} collisions are shown and compared to results for the same set of calculations (see text).} 
\label{fig:hypertritonoverLambda}
\end{figure}

\begin{figure}[htbp]
	\centering 
	\includegraphics[width=\linewidth]{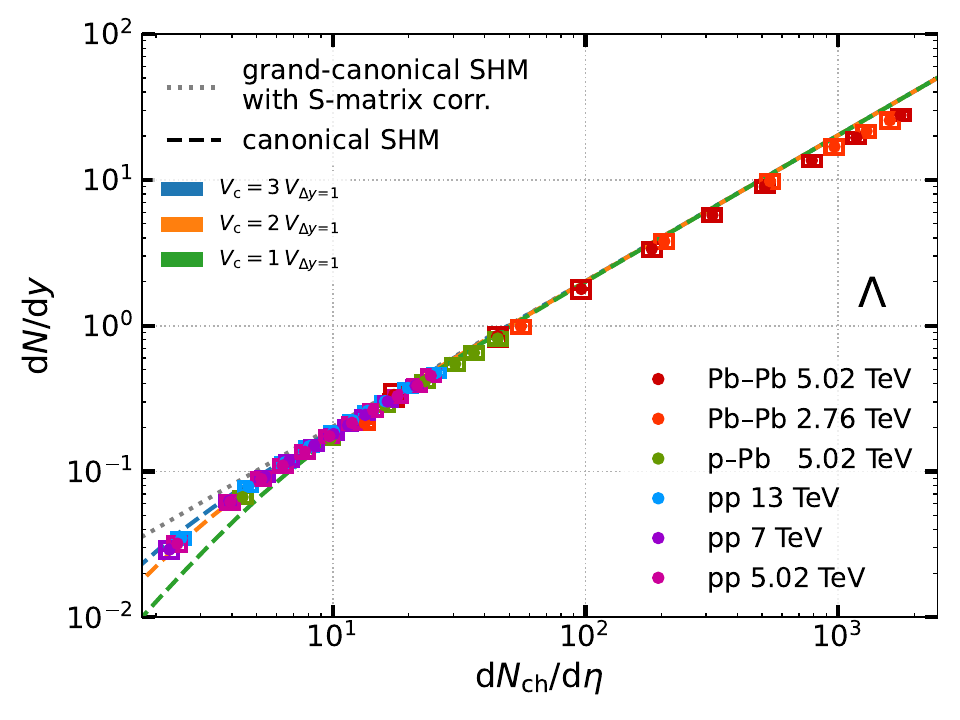}	
	\caption{Multiplicity dependence of $\Lambda$ rapidity densities. The same line styles as in Fig.~\ref{fig:deuteron} are used. The S-matrix correction is applied to the SHM yield~\cite{Cleymans:2020fsc}. ALICE measurements for Pb--Pb~\cite{ALICE:2025cqy,ALICE:2013cdo}, p--Pb~\cite{ALICE:2013wgn}, and pp~\cite{ALICE:2019avo,ALICE:2016fzo} collisions are shown. Error bars (open boxes) represent statistical (systematic) uncertainties.} 
	\label{fig:suppLambda}
\end{figure}

\section*{Acknowledgments}
We thank Andreas Nogga for providing the one-body densities for the (hyper)nuclei studied in this paper. We thank Benjamin Dönigus and Hans-Werner Hammer for stimulating discussions.
\bibliographystyle{elsarticle-num} 
\bibliography{bibliography_nucleipaper}
\end{document}